\documentclass[aps,preprint,preprintnumbers,nofootinbib,showpacs]{revtex4}

\def\beq{\begin{equation}}
\def\eeq{\end{equation}}
\def\beqa{\begin{eqnarray}}
\def\eeqa{\end{eqnarray}}

\def\ssc{\scriptscriptstyle}
\def\lsim{\mathrel{\raise.3ex\hbox{$<$\kern-.75em\lower1ex\hbox{$\sim$}}} }
\def\gsim{\mathrel{\raise.3ex\hbox{$>$\kern-.75em\lower1ex\hbox{$\sim$}}} }

\begin{document}
\draft
\preprint{\tighten{\vbox{\hbox{NCU-HEP-k036}
\hbox{rev. Sep 2010}
}}}
\vspace*{.5in}

\title{Poincar\'e-Snyder Relativity with Quantization
\vspace*{.3in} }
\author{\bf Otto C. W. Kong and Hung-Yi Lee\vspace*{.2in}}
\email{otto@phy.ncu.edu.tw}
\affiliation{
Department of Physics and Center for Mathematics and Theoretical Physics,
National Central University,~Chung-li,~TAIWAN 32054.
 	}

\vspace*{.5in}
\begin{abstract} 
Based on a linear realization formulation of a quantum relativity, 
--- the proposed relativity for quantum `space-time', 
we introduce the new Poincar\'e-Snyder relativity
and Snyder relativity as relativities in between the latter and the well known
Galilean and Einstein cases. We discuss how  the Poincar\'e-Snyder relativity may
provide a stronger framework for the description of the usual (Einstein) relativistic 
quantum mechanics and beyond with particular focus first on a geometric quantization
picture through the $U(1)$ central extension of the relativity group, which
had been establish to work well for the Galilean case but not for the Einstein case.
\end{abstract}
\pacs{02.90.+p,03.65.-w, 03.65.Pm, 03.30.+p}

\maketitle

\section{Introduction}
Possible relativities as described by relativity symmetries beyond the
Lorentz or Poincar\'e group of Einstein relativity have been catching
quite some interest recently (see for example Refs.\cite{dsr,G}).
Since the pioneering work of Snyder\cite{S}, symmetry deformation, mostly
considered as required to implement an invariant quantum scale, has
been a main key for the direction of theoretical pursuit. That gives the
idea of a quantum relativity. On the other hand, if one does believe that
the entity we used to know as space-time does have a different structure 
at the true microscopic/quantum level that can plausibly be described 
directly, such a `quantum space-time' will have its own relativity.
The relativity symmetry deformations could be nicely formulated as
Lie algebra stabilizations \cite{CO}. Following the line of thinking,
we implemented in Ref.\cite{023} a linear realization perspective to arrive
at the `quantum space-time' description with the quantum relativity
symmetry as the starting point.  Lorentz or Poincar\'e symmetry (of Einstein
relativity) can be considered exactly a result of the stabilization of the
Galilean relativity symmetry. The linear realization scheme in that setting is 
nothing other than the Minkowski space-time picture. Such a mathematically 
conservative approach, however, leads to a very radical physics perspective
that at the quantum level space-time is to be described as part of something 
bigger \cite{023}. The latter as the arena for the description of the new
fundamental physics is called the quantum world 
in Ref.\cite{030}. It is to be identified, mathematically, as the coset
space $SO(2,4)/SO(2,3)$ \cite{031}, or the hypersurface
$\eta_{\ssc \mathcal M\mathcal N} X^{\!\ssc \mathcal M} X^{\!\ssc \mathcal N} = 1$
[$\eta_{\ssc \mathcal M\mathcal N} =( -1, 1, 1, 1, 1, -1)$],
\footnote{Note that we have flipped the metric sign convention adopted
in our earlier publications \cite{023,030,031}; from now on, the
time-like (space-like) geometric signature is -1 (+1)}
within the 6D classical geometry with $X^\mu$ ($\mu=0$ to 3) being space-time
coordinates while $X^4$ and $X^5$ being {\it non-space-time} ones. The `time' of
Minkowski space-time is not just an extra spatial dimension. Its nature is
dictated, from the symmetry stabilization perspective, by the physics of having
the invariant speed of light $c$. The other two new coordinates in our `quantum 
space-time' picture are likewise dictated to be neither space nor time\cite{023,030}.
We reproduce in table~1 the the suggested physics of the stabilizations/deformations 
involved in our stabilizations and extensions by translations (of the corresponding 
arenas for the linear realizations) sequence arriving at the $SO(2,4)$ quantum 
relativity \cite{030} as illustrated by 
{\boldmath \beqa \nonumber
&& ISO(3) \; \rightarrow \;\; SO(1,3)\;\;\hookrightarrow\;\; ISO(1,3) \\
&&\;\; \rightarrow \; SO(1,4)
 \;\; \hookrightarrow \;\; ISO(1,4)  \; \rightarrow \;\; SO(2,4) \;.
\nonumber \eeqa}
Like $X^0=ct$, we have $X^4=\kappa c \sigma$ and $X^5=\ell\rho$ with
however, $\sigma$ having the physics dimension of time/mass (and a space-like
geometric signature) and $\rho$ a pure number. Understanding the physics role of the 
two extra coordinates $\sigma$ and $\rho$ of the quantum world is considered crucial
for any attempt to formulate the dynamics. Here in this letter, we report
a way to approach the challenge --- analyzing the physics of what
we called the Poincar\'e-Snyder relativity.

We will explain first, in the next section, structure of the Poincar\'e-Snyder
relativity, with symmetry denoted by $G(1,3)$. In short, it is mathematically sort 
of a `Galilie group' for 4D space-time. The analog of time $t$ as an external
evolution parameter for Galilean dynamics is here given by $\sigma$.
Recall that $\sigma$ has a space-like geometric signature but the physics
dimension of time/mass\cite{023,030}. We are inspired to consider the new
relativity by our studies on the quantum relativity. The ultimate goal is to
analyze and formulate physics directly for the intrinsically quantum arena ---
the quantum world (see discussions in Refs.\cite{030,031}). To better
prepare ourselves for the formidable challenge, we want to take a step backward
and study the relativity(ies) with symmetry between the Einstein and our
quantum case. From the latter perspective,  the Poincar\'e-Snyder
relativity is the first step beyond Einstein relativity. Its physics setting
should be not much different from the latter. It has, however, a
mathematical structure very similar to the $G(3)$ Galilean case. The latter
suggests similar mathematics in the formulation of the admissible dynamics for
the two cases, both at the classical and quantum level. The Poincar\'e-Snyder
relativity mechanics may hence be a much more familiar object. We must warn
the readers that the physics interpretations of the similar mathematics
are expected to be quite nontrivial and un-conventional though. 

The Poincar\'e-Snyder relativity is still a relativity on 4D Minkowski
space-time, only with an extra kind of momentum
dependent reference frame transformations admitted. These momentum boosts
are independent of the usual velocity boosts, but reduce to the latter
when $\sigma=\tau/m$, the Einstein proper time over the standard (fixed)
particle rest mass \cite{023}. Just as Galilean velocity boosts are
transformations on 3D space dependent on an external parameter time,
the momentum boosts enforce the independent $\sigma$-coordinate external
to 4D space-time. The `dynamic' formulation naturally suggests taking
$\sigma$ as a sort of `evolution' parameter. We called that $\sigma$-dynamics
or $\sigma$-evolution, withholding the exact physics interpretation.
Within the Einstein framework, $\sigma$-evolution looks like
proper time evolution, and as such have actually been used quite a lot in
the literature to describe Einstein relativistic dynamics, both classical
and quantum\cite{HP,cS}. This letter is the first step to take a second
look at the kind of studies, focusing on the difference and superiority
of the new Poincar\'e-Snyder perspective. In particular, we will present
in section 3 results of the picture of quantization as $U(1)$ central
extension \cite{book}. Notice that unlike the quantum relativity, and
possibly the Snyder relativity obtained from the stabilization of the
Poincar\'e-Snyder relativity, the construction of the $G(1,3)$ symmetry
involves none of the quantum physics inspired deformations with quantum scale(s)
as deformation parameters. Hence, there is no reason at all to expect the
relativity to be in any sense quantum. It looks only like a different
perspective to look at classical physics on Minkowski space-time; and as such
should be liable to quantization. Results of section 3 actually lay further
justification to that {\em a posteriori}.
Of course the ultimate justification for the $G(1,3)$
approach from the theoretical side has to come from the relativities
beyond. Or there is the possibility of seeing experimental evidence for
Poincar\'e-Snyder relativity or Snyder relativity through careful studies of
the $\sigma$-dynamics and its physics interpretations.

The kind of physics picture we have in mind
behind our work and the earlier papers \cite{023,030} is not quite like any
of the familiar old pictures, and admittedly not yet fully conceptuallized. 
The research program is a very ambitious one, aiming at dynamics beyond any
existing framework. We find the need to take the most conservative strategy,
trying to commit to the minimal conceptual physics picture on any particular
new aspects arising from the formulation before one can be sure that it is the right
way to look at it. We try to learn from the mathematics and logics of the basic
formulation what it can offer. One will see that such a conservative strategy
can still bring out quite some interesting results presented in this letter and in 
another accompanying long paper \cite{037}.

{\it Readers be very cautious.}
This letter can be read without detailed reading of the earlier papers, but what
motivates a particular definition or approach would then be difficult to appreciate.
With or without reference to Refs.\cite{023,030}, it is
important for readers to read first what we presented as it is, without assuming
a perspective from any other theory standard or less conventional. This is
especially true with the very similar looking structure from the line of
work on covariant formulation of Einstein relativistic dynamics\cite{HP,cS}.
We highlight here a few crucial points to bear in mind.  
The momentum boosts are newly introduced transformations with physical meaning 
still to be clarified (see \cite{023} for some discussions). That comes with a 
modified or sort of generalized definition of energy-momentum, as $\sigma$ 
derivatives. The $\sigma$ parameter is of central importance, with physics 
content still to be fully understood. It is definitely not a measure of time. 
We will introduce the dynamics of the Poincar\'e-Synder relativity as a formal 
$\sigma$-evolution, more or 
less duplicating mathematically the time dynamics of the Galiliean case. In fact,
a main aim of the studies is to learn about how to think about the physics
of the $\sigma$ parameter. Comparing any expression here with similar ones
having essentially a time evolution pespective from the physics point of 
view leads only to confusion. In fact, in the long paper to follow, we will
illustrate the right way to really look at the time evolution results our
$\sigma$-evolution formulation provides in a more fundamental setting --- that
of canonical classical mechanics. For instance, we have derived directly from
the formulation an interesting solution of particle-antiparticle annihilation,
which is considered to be a nontrivial success of our approach \cite{037}.


\section{Poincar\'e-Snyder Relativity}
The Poincar\'e-Snyder relativity is a relativity on 4D Minkowski space-time, 
with $\sigma$ as an external parameter. It has otherwise a structure
mimic that of the Galilean case on 3D space. The complete Galilei 
group has rotations, translations, as well as velocity boosts as symmetry 
transformations on 3D space together with an external time parameter.
Comparing the first two columns of table~1, we can see that the
implementation of Poincar\'e symmetry stabilization through the
invariant (quantum) energy-momentum scale requires considering
a new kind of momentum boosts, as independent from the velocity boosts.
The usual relation between momentum and velocity has to be relaxed
to hold only at the Einstein limit\cite{023}.
The Poincar\'e-Snyder relativity is then just the relativity with
Poincar\'e symmetry extended by such momentum boosts before the deformation,
{\it i.e.} at the unconstrained commuting limit. The relativity  may hence
provide a window for us to understand the $\sigma$ parameter in the most
familiar context. Interestingly enough, we came to realize that
parameter(s) of quite close a nature to that of $\sigma$ had been used quite
a lot, to different extents, in the regime of (Einstein) relativistic
quantum mechanics in somewhat ambiguous ways \cite{HP,cS}.
Adopting the perspective of the Poincar\'e-Snyder relativity we
propose here actually helps to put many of such earlier attempts
on solid theoretical footings. From the perspective alone, that comes 
at a great cost though --- a new definition of energy-momentum as $\sigma$ 
derivatives with a physics picture still to be fully understood \cite{023}. 

Let us start by a clear illustration
of the structure of our Poincar\'e-Snyder relativity.
Following Ref.\cite{Gilmore} (see Fig.~10.6 for an illustration), we
can describe the Poincar\'e group and the Galilei group as sequential
contractions from $SO(1,4)$
\[ SO(1,4) \longrightarrow ISO(1,3) \longrightarrow G(3) \;. \]
The first step is the well known In\"on\"u-Wigner contraction, a reverse
of the symmetry stabilization. A further, similar, contraction gives the
Galilei group with commuting translations as well as commuting velocity boosts.
We are more interested in the other contraction sequence, from the symmetry
of our quantum relativity. That is the sequence
\[ SO(2,4) \longrightarrow ISO(1,4) \longrightarrow G(1,3) \;, \]
giving the newly named Snyder and Poincar\'e-Snyder relativities. In table~2, we 
list the full set of generators for the symmetry groups of all the five relativities. 
\footnote{
In a forthcoming paper, we will present more details of the mathematics and
plausible physics pictures of relativity symmetries from the contraction schemes
based on the kind of perspective \cite{041}.}
Wherever there is a change of notation from the
$J_{\ssc \mathcal M\mathcal N}$ generator(s) moving across a row,
a contraction is involved. $J_{\ssc \mathcal M\mathcal N}$ here denotes
the 15 generators of the the {\boldmath\small $so(2,4)$} algebra, satisfying
\beq \label{so}
[J_{\ssc \!\mathcal R\mathcal S}, J_{\ssc\! \mathcal
M\mathcal N}] = - 
( \eta_{\ssc \mathcal S\mathcal M} J_{\ssc
\mathcal R\mathcal N} - \eta_{\ssc \mathcal R\mathcal M} J_{\ssc
\mathcal S\mathcal N} + \eta_{\ssc \mathcal R\mathcal N} J_{\ssc
\mathcal S\mathcal M} -\eta_{\ssc \mathcal S\mathcal N} J_{\ssc
\mathcal R\mathcal M}) \;, 
\eeq
where again $\eta_{\ssc \mathcal M\mathcal N} =( -1, +1,
+1, +1, +1, -1)$ with the indices go from $0$ to $5$; we use also
$\eta_{\ssc A\!B}$ to denote the $0$ to $4$ part; others
indices follow the common notation . Note that the
$J_{\mu\nu}$'s are the (subset of) Lorentz transformation generators, etc.
All $P_{\ssc A}$'s denote (generators for) the coordinate translations on
the corresponding arena for the linear realizations
--- $M^5$, the Minkowski space-time $M^4$, and the 3D space. The $K_i$'s are
Galilean velocity boosts, and $K_\mu^{\prime}$'s the new commuting momentum boosts.
We have the standard structure
\beq
[J_{\ssc \!A\!B}, P_{\ssc C}] = -
(\eta_{\ssc B\!C} P_{\ssc A}
- \eta_{\ssc A\!C} P_{\ssc B}) \;,
\eeq
and
\beq
[J_{\mu\nu}, K_\lambda^{(\prime)}] = -
 (\eta_{\nu\lambda} K_\mu^{(\prime)}
- \eta_{\mu\lambda} K_\nu^{(\prime)} ) \;,
\eeq
where the latter applies to the two different types of boosts $K_\mu^{\prime}$
and $K_i$'s. Translations and the boosts (not including the so-called Lorentz
boosts which are space-time rotations) always from a commuting set. The external
time evolution $H$ commutes with all others with the only exception given by
\beq
[K_i, H] = 
 P_i \;.
\eeq
The latter is actually the same commutation relation as that of the generators
$J_{0i}$ and $P_{\ssc 0}$ before the {$ISO(1,3) \longrightarrow G(3)$}
symmetry contraction. In fact, no commutation relation between time translation
and the other generators is changed in the contraction. We choose to use
$H$ instead of $P_{\ssc 0}$ to denote time translation for the Galilean case
only to highlight time being a parameter external to the geometric realization arena
of 3D space. Similarly the external $\sigma$-translation $H^{\prime}$ has the only
non-zero commutators
\beq
 [K_\mu^{\prime}, H^{\prime}] = 
P_\mu \;,
\eeq
equal to that of the corresponding ones between $J_{{\ssc 4}\mu}$ and $P_{\ssc 4}$.

Note that the translations for Einstein and Galilean relativities are listed
in the rows of the  $J_{{\ssc 4}\mu}$'s and $K_\mu^{\prime}$'s for the other
three cases, rather than with the other $P_\mu$ translations. That is done to
emphasize that all the 10 generators of the Poincar\'e or Galilei group,
like the case of the $J_{\mu\nu}$ and $K_\mu^{\prime}$ subset for the $G(1,3)$
group of the Poincar\'e-Snyder relativity, can be obtained through proper contractions
of an $SO(1,4)$ symmetry. This is more in correspondence with the stabilization 
sequence presentation in our earlier publications \cite{023,030}, in which we make 
no clear distinction nor simultaneous treatment of boosts and translations. 
The $J_{\mu\nu}$ and $K_\mu^{\prime}$ set indeed gives an algebra isomorphic to 
that of the $J_{\mu\nu}$ and $P_\mu$ set. However, one can also
keep all the $P_i$'s and $P_{\ssc 0}$ on the same role. That corresponds to
seeing the last two groups as from the equally valid contraction sequence
\cite{Gilmore}
\[ SO(2,3) \longrightarrow ISO(1,3) \longrightarrow G(3) \;, \]
perhaps more adapted to tracing back their relations to the full $SO(2,4)$
symmetry. We are interested here mostly in illustrating the structure of the
$G(1,3)$ symmetry for the Poincar\'e-Snyder relativity.

As suggested by the notation, the $G(1,3)$ symmetry has a very similar
structure to that of the Galilean $G(3)$, hence similar mathematical
properties. The latter may imply similar properties, at the level of 
mathematical formulations, when applied to describe physics.
Two main features are considered specially interesting.
Firstly, taking away only $H$ from the set of generators of $G(3)$,
the rest generates a subgroup, likewise for taking away $H^{\prime}$
in the case of $G(1,3)$. This is not the case for the set of $ISO(1,3)$
generators with $P_{\ssc 0}$, for example. Particle dynamics under
Poincar\'e symmetry has a no-interaction theorem \cite{no-int,Ste}. The latter
can be interpreted as a consequence of different subgroup structure,
as compared to that of the $G(3)$ symmetry. All generators besides the Hamiltonian
$H$ for the symmetry stay as kinematical ones, which has to generate a subgroup.
The generators besides the Hamiltonian fail to do the same for Poincar\'e symmetry,
leaving rather the three admissible forms of Hamiltonian dynamics as noted by 
Dirac\cite{D} (see also Ref.\cite{Ste}) as the alternatives. The generators besides 
the $H^{\prime}$ Hamiltonian of the $G(1,3)$ symmetry can be taken as all being
kinematical.  Using $G(1,3)$ symmetry to describe `relativistic dynamics', or rather
the dynamics of $\sigma$-evolution on Minkowski space-time, would admit direct 
description of interactions as in the Galilean case. It will be interesting to see
if we can learn something about `relativistic dynamics' from such an
approach (see Ref.\cite{037}).  Next, we turn to a feature that we want
to focus on here. The $G(1,3)$ group, like $G(3)$, admits a non-trivial
$U(1)$ central extension. Projective group representations required
to describe quantum mechanics are simply unitary representations
of the $U(1)$ central extension \cite{book}. Hence, the $G(1,3)$ may be
a better candidate than the $ISO(1,3)$ for the description of
`relativistic quantum mechanics' as a quantization of `relativistic
mechanics'.

\section{\boldmath Quantization  as $U(1)$ central extension}
Aldaya and de~Azc\'arraga \cite{gq} presented a particularly nice approach to
geometric quantization in which the quantum dynamical description of the system
can be obtained with the symmetry group as the basic starting point (see also
Ref.\cite{book}). The approach looks especially relevant to our case in
which we have a new relativity symmetry in search of a clear understanding
of the physics involved. In fact, while the approach gives an elegant
presentation for the quantization of the Galilean system, its application
to the case of Einstein relativity is less than equally appealing. For
the former case, the group to be considered is a $U(1)$ central extension
of the symmetry for the corresponding classical system --- $G(3)$. 
The essentially unique nontrivial central extension is
depicted by the modified commutator $[K_i, P_j]= m \delta_{ij} \Xi$, where
$\Xi$ is a central charge ($m$ the particle mass). The $ISO(1,3)$ symmetry,
however, admits no such nontrivial central extension (for an explicit
discussion on admissible central charges for both cases, see Ref.\cite{Ste}).
It is easy to see from our discussion above of the $G(1,3)$ symmetry for the
Poincar\'e-Snyder case that it has  a structure mostly parallel to that
of the Galilean $G(3)$. Hence, we should have the same nontrivial
$U(1)$ central extension available for the implementation of such a
quantization scheme. Indeed, when we set out to perform the analysis, we
realized that the work, at least most of the mathematics, has actually
been done \cite{pkg} under a different premise. Confronted with the
difficulty on applying the elegant quantization scheme to (Einstein)
relativistic dynamics, Aldaya and de~Azc\'arraga chose to put the
Poincar\'e symmetry into a mathematical framework that make the scheme
applicable --- essentially taking it to $G(1,3)$. They basically considered
promoting the proper time to an `absolute time' parameter for the formulation.
The latter was used more like as a mathematical trick with any independent
physics meaning not explicitly addressed. The physics results are discussed
only after a symmetry reduction back to the
Einstein setting has been implemented (see also Ref.\cite{ap}).
We choose here to follow mostly the approach of Ref.\cite{gq} and
present first the quantization results in the language of our Poincar\'e-Snyder
relativity formulation. After that we will discuss the very important difference 
in physics premise and interpretation we introduce here. We discuss why the
Poincar\'e-Snyder relativity perspective is considered to provide a plausibly more
interesting framework for the bold attempt at the group quantization
formulation of the `quantum relativistic system'. Our approach may
also provides an interesting way to avoid the many `uncomfortable' features
well appreciated in the usual (Einstein) relativistic quantum mechanics, which
otherwise need to be resolved in the quantum field theory framework.

The standard action of $G(1,3)$ on the Minkowski spacetime ($x^\mu$)
with the extra, external, parameter $\sigma$ is given by
\beqa
&& x'^\mu=  {\Lambda^{\mu}}_{\!\nu} x^\nu + p^\mu \sigma + A^\mu \;,
\nonumber \\
&& \sigma' = \sigma + b \;.
\eeqa
An element of our extended $G(1,3)$ group may be written as
$g=(b, A^{\mu}, p^{\mu}, {\Lambda^{\mu}}_{\!\nu}, e^{i\theta})$,
with group product rule given by 
\beqa
&& A''^\mu=  {\Lambda'^{\mu}}_{\!\nu} A^\nu + p'^\mu b + A'^\mu \;,
\qquad
b'' = b' + b \;,
\nonumber \\
&& p''^\mu = {\Lambda'^{\mu}}_{\!\nu} p^\nu + p'^\mu \;,
\qquad\qquad
{\Lambda''^{\mu}}_{\!\nu} ={\Lambda'^{\mu}}_{\!\rho}{\Lambda^{\rho}}_{\!\nu}\;,
\eeqa
and the nontrivial $U(1)$ extension of given by
\beq
\theta''=\theta' +\theta + z \left[A'_\mu  {{\Lambda'^{\mu}}_{\!\nu}} p^\nu
+ b(p'^\mu {{\Lambda'_{\mu}}^{\!\nu}} p_\nu
+\frac{1}{2} p'^\mu p'_\mu)\right] \;.
\eeq
The last term is the cocycle the exact choice of which is
arbitrary up to a coboundary \cite{book}; $z$ corresponds to a value
of the central charge is taken as an arbitrary constant at this point.

The right-invariant vector fields are given by
\beqa
&& {\bf X}^{\!\!\ssc R}_b = \frac{\partial}{\partial b} \;,
\qquad \qquad
{\bf X}^{\!\!\ssc R}_{A^\mu} = \frac{\partial}{\partial A^\mu}
+ z p_ \mu \; \frac{i\zeta}{\hbar} \frac{\partial}{\partial \zeta} \;,
\nonumber \\
&& {\bf X}^{\!\!\ssc R}_{p^\mu} = b\frac{\partial}{\partial A^\mu}
+ \frac{\partial}{\partial p^\mu} + z b p_\mu \;
\frac{i\zeta}{\hbar} \frac{\partial}{\partial \zeta} \;,
\nonumber \\
&& {\bf X}^{\!\!\ssc R}_{\omega^{\mu\nu}}
 =  {\bf \widetilde{X}}^{\!\!\ssc R}_{\omega^{\mu\nu}}
+ A_\nu \frac{\partial}{\partial A^\mu}- A_\mu \frac{\partial}{\partial A^\nu}
+ p_\nu\frac{\partial}{\partial p^\mu}- p_\mu\frac{\partial}{\partial p^\nu}\;,
\nonumber \\
&& {\bf X}^{\!\!\ssc R}_\zeta = \frac{i\zeta}{\hbar} \frac{\partial}{\partial \zeta} \;,
\eeqa
where we skip the details of ${\bf \widetilde{X}}^{\!\!\ssc R}_{\omega^{\mu\nu}}$, the invariant
vector field for the $SO(1,3)$ subgroup
[with $\Lambda(\omega) =e^{\frac{-i}{2}\omega^{\mu\nu} J_{\mu\nu}}$],
leaving it to be given explicitly in the appendix.
Note that $\zeta=\exp({\frac{i}{\hbar}\theta})$ with
$\frac{i\zeta}{\hbar} \frac{\partial}{\partial \zeta}= \frac{\partial}{\partial \theta}$
locally. The left-invariant vector fields are given by
\beqa
&& {\bf X}^{\!\!\ssc L}_b = \frac{\partial}{\partial b}
+ p^\mu \frac{\partial}{\partial A^\mu}
 +\frac{z}{2}p^{\mu}p_{\mu} \; \frac{i\zeta}{\hbar} \frac{\partial}{\partial \zeta}\;,
\nonumber \\
&&
{\bf X}^{\!\!\ssc L}_{A^\mu} =
\frac{\partial}{\partial A^\nu}{{\Lambda^{\nu}}_{\!\mu}}\;,
\qquad \qquad
{\bf X}^{\!\!\ssc L}_{p^\mu} =  \frac{\partial}{\partial p^\nu} {{\Lambda^{\nu}}_{\!\mu}}
+ z A_\nu  {{\Lambda^{\nu}}_{\!\mu}} \;\frac{i\zeta}{\hbar} \frac{\partial}{\partial \zeta}\;,
\nonumber \\
&&  {\bf X}^{\!\!\ssc L}_{\omega^{\mu\nu}}
 =  {\bf \widetilde{X}}^{\!\!\ssc L}_{\omega^{\mu\nu}} \;,
\qquad \qquad
{\bf X}^{\!\!\ssc L}_\zeta = \frac{i\zeta}{\hbar} \frac{\partial}{\partial \zeta} \;;
\eeqa
again explicit expression for the $SO(1,3)$ vector field
${\bf \widetilde{X}}^{\!\!\ssc L}_{\omega^{\mu\nu}}$ is left to the appendix.
We have the quantization form given by the left-invariant 1-forms conjugate
to ${\bf X}^{\!\!\ssc L}_\zeta$, the vertical 1-form; explicitly
\beq
\Theta
= -z A^\nu{{\Lambda_{\nu}}^{\!\mu}} dp_\mu  - \frac{z}{2}p^{\mu}p_{\mu} db
+  \frac{\hbar d\zeta}{i\zeta} \;.
\eeq
The characteristic module is defined through the conditions
$i_{\bf X}\Theta=0$ and $i_{\bf X}d\Theta=0$, characterizing the
differential system given by the vector field ${\bf X}^{\!\!\ssc L}_b$.
We have the equations of motion given by
\beqa
&& \frac{db}{d\sigma}=1 \;,
\qquad \qquad
\frac{dA^\mu}{d\sigma}=p^\mu \;,
\nonumber \\
&&
\frac{dp^\mu}{d\sigma}=0 \;,
\qquad \qquad
\frac{d{{\Lambda^{\mu}}_{\!\nu}} }{d\sigma}=0 
\quad \left(\;\frac{d{{\omega}^{\mu\nu}} }{d\sigma}=0\;\right) \;,
\nonumber \\
&& \frac{d\zeta}{d\sigma}= \frac{z}{2}p^{\mu}p_{\mu} \frac{i\zeta}{\hbar} \;.
\eeqa
Identifying the integration parameter as $\sigma$ gives
\beqa
&& b=\sigma \;,
\qquad \qquad
A^\mu = p^\mu \sigma + K'^\nu {\Lambda_{\nu}}^{\!\mu}\;,
\qquad \qquad
p^\mu=P^\mu \;,
\nonumber \\
&& \zeta= \zeta_{o} \exp(\frac{iz}{2\hbar}p^{\mu}p_{\mu}\sigma)\;.
\eeqa
Naturally, $A^\mu$ is to be identified as $x^\mu$ giving $p^\mu$ 
as $\frac{dx^\mu}{d\sigma}$, showing consistence with our original introduction
of the momentum boosts (see table 1) as the extra symmetry transformations
to supplement $SO(1,3)$ and hence getting to $G(1,3)$. We have constants
of motion, $K'^\nu {\Lambda_{\nu}}^{\!\mu}$, $P^\mu$, and
$\zeta_{o}$ which parametrize the quantum manifold. Passing to the latter,
$\Theta$ takes the form
\beq
\Theta_{\!P} = - z K^\mu dP_\mu +  \frac{\hbar d\zeta_o}{i\zeta_o} \;.
\eeq

The symplectic 2-form is given by $\omega =d \Theta$. Taking $z=1$,
$H'=\frac{1}{2}p^{\mu}p_{\mu}$, and $K'^\mu=A^\nu{{\Lambda_{\nu}}^{\!\mu}}$
we have
\beq
\omega=d\Theta_{\!P} = - dK'^{\mu}\wedge dp_\mu 
\;,
\eeq
where we have taken $z=1$ corresponding to $H'=\frac{1}{2}p^{\mu}p_{\mu}$,
which gives the right form for the classical $\sigma$-Hamiltonian \cite{037}.
The expression suggested the identification of $(K'^\mu,b)$ as
particle `position' variables $(x^\mu,\sigma)$ and $H'$ as the
$\sigma$-Hamiltonian generating `evolution' in the absolute parameter $\sigma$.
The prequantum operator associated with a real function $f$ on
the classical phase space acting on wavefunction $\psi$ is given by
\beq
\hat{f} \psi \equiv  - i\hbar \tilde{X}f\cdot \psi
=-i\hbar X_f \cdot \psi + [f- \Theta(X_f)] \psi \;,
\eeq
where  $i_{X_f} \omega= -df$. In particular,
\beqa
&&\hat{K'}^\mu = i\hbar \frac{\partial}{\partial P_\mu}  \;,
\qquad\qquad
\hat{P}_\mu= -i\hbar  \frac{\partial}{\partial K'^\mu} + P_\mu\;,
\nonumber \\
&&\hat{\sigma}= i \hbar\frac{\partial}{\partial H'}  \sigma \;,
\qquad\qquad
\hat{H'}= i \hbar \frac{\partial}{\partial \sigma} \;,
\eeqa
where the an extra negative sign is adopted in $\hat{H}'$.
The operators  $K'^\mu$ and $P_\mu$ can also be obtained from
${\bf X}^{\!\!\ssc R}_{A^\mu}$ and ${\bf X}^{\!\!\ssc R}_{p^\mu}$.
The full polarization subalgebra can be taken as spanned by
$\{{\bf X}^{\!\!\ssc L}_b,
{\bf X}^{\!\!\ssc L}_{A^\mu}, {\bf X}^{\!\!\ssc L}_{\omega^{\mu\nu}}\}$,
giving the momentum space wavefunction $\phi(P_\mu)$, {\it i.e.} the wavefunction
as dependent only on $P_\mu$ but not $K'^\mu$. 
We have then simply $\hat{P}_\mu \phi(P_\mu)= {P}_\mu \phi(P_\mu)$.
${\bf X}^{\!\!\ssc L}_b$ generates the Euler-Lagrange equation
\beq \label{covS}
i \hbar\partial_\sigma \psi +  \frac{\hbar^2}{2}\partial_\mu \partial^\mu \psi = 0
\eeq
for the Fourier transform $\psi$ of `momentum' space wavefunction $\phi$.
Note that $\hat{H}'$ and $\hat{P}_\mu$ constitute a complete set of commuting
observables for the configuration space wavefunction.
The equation expresses an operator form of
$H'=\frac{1}{2}p^{\mu}p_{\mu}$ with $\hat{P}_\mu$ reduced to just
$-i\hbar  \frac{\partial}{\partial K'^\mu}$, {\it i.e.}
$-i\hbar  \frac{\partial}{\partial x'^\mu}$. Eq.(\ref{covS}) is of the same form
as the so-called (Lorentz) covariant Schr\"odinger equation studied in the
literature\cite{cS}, except with the parameter $\sigma$ replacing the proper time
$\tau$ (or rather $\tau/m$). The equation, with again essentially the proper time as 
evolution parameter, is also what is obtained in Ref.\cite{pkg}. One can see that the 
rest mass, or $m^2/2$ to be exact, of an Einstein particle is just the $\hat{H}'$ 
eigenvalue. Without considering the spin degree of freedom, the usual (Einstein) 
relativistic quantum mechanics corresponds to the $\sigma$ independent eigenvalue equation
for $\hat{H}'$, obtainable from Eq.(\ref{covS}) separation of the $\sigma$ variable
from the $x^\mu$ variables. The eigenvalue equation is the Klein-Gordon equation.
Recall that for an Einstein particle, {\it i.e.} taking the Poincar\'e-Snyder free
particle to the Einstein relativistic limit, we have $\sigma \to \tau/m$ \cite{023,030}.

\section{Discussions}
One can see from the above that Poincar\'e-Snyder relativity  provides a
very nice mathematical framework to formulate the the covariant quantum
mechanics except with the Lorentz invariant evolution parameter $\tau$
replaced by the truly independent variable $\sigma$ as suggested from the
quantum relativity framework. The introduction of an extra evolution
parameter in the beautiful quantization scheme of Ref.\cite{pkg} and the various
early discussions of the covariant Schr\"odinger before the 50's~\cite{cS}
as sort of a mathematical tool is now dictated by the quantum relativity
picture. It remains a challenge to interpret the $\sigma$ dependent mechanics
at both the quantum and the classical level beyond the case of an Einstein particle.
We emphasize again that $\sigma$ is not a kind of time parameter. 
The key lesson from our perspective is that one has to go beyond thinking about
the `evolution' parameter as essentially time, which confines all earlier literature.  
In a somewhat different background
setting, a first discussion of the physics of the $\sigma$ coordinate has been
given in Ref.\cite{023}. The most important point to note is that the framework
actually redefine energy-momentum through $p^\mu=\frac{dx^\mu}{d\sigma}$,
making it not equal to the Einstein form of $m \frac{dx^\mu}{d\tau}$.
In general, for the quantum relativity or the Poincar\'e-Snyder relativity,
particle rest mass becomes a reference frame dependent quantity. A momentum
boost transformation changes the value of $m$ as the magnitude of the
energy-momentum four-vector. In Poincar\'e-Snyder relativity, the vector
transforms by simple addition like the Galilean velocity. This is the new
and most fundamental feature offered by our framework. A related aspects is the lost
of the Einstein rest mass as an intrinsic or fundamental character of a particle.
Here, it is only the magnitude of the particle energy-momentum four-vector
which can be modified by interaction. The feature is illustrated to be useful,
or even necessary, in describing some interesting physics scenario like
particle-antiparticle annihilation\cite{037}.

In Galilean relativity, the kinetic energy of a particle ($\propto v^2$) is
both reference frame dependent and interaction dependent hence time dependent.
Similarly, the (expectation) value of the $\sigma$-Hamiltonian is, in general,
$\sigma$ dependent. To put it another way, the $\sigma$ dependent covariant
Schr\"odinger equation is to be given by
\beq
i \hbar\partial_\sigma \psi -  \hat{H}' \psi = 0
\eeq
where $\sigma$-Hamiltonian $\hat{H}'$ operator should be given by
$\frac{1}{2}\hat{P}^\mu\hat{P}_\mu + \hat{V}'$ with $V'$ depicting an `interaction
potential' under the  $\sigma$-evolution. The value for the magnitude of the
energy-momentum four-vector would hence change with $\sigma$. Such a picture
is fully collaborated by a classical canonical Hamiltonian picture\cite{037}.

It is interested to note that in various studies of the essentially
$\tau$-parametrized covariant Schr\"odinger equation there had been discussions
of notion of mass indefiniteness~\cite{HP,cS}. Naively, Eq.(\ref{covS}) admits
mixture of eigenstates of different $m^2$ value. Some author actually went
so far as to absorb $m$ into the evolution parameter and made it $\tau/m$,
which is indeed close to our $\sigma$. An example of an explicit physics
considerations of statistical nature, for example, was offered by Hostler\cite{H}.
In our opinion, Feynman was the one that went beyond everybody, in his works on
quantum electrodynamics. Not only did he rewrote the Klein-Gordon equation
in the form of Eq.(\ref{covS}) with $u\equiv\tau/m$ in the place of $\sigma$ \cite{F},
he actually considered the case of $dt/du<0$ hence taking the `evolution' somewhat
beyond a physical time variable. That was actually behind the antiparticle picture
in Feynman diagrams used in the standard quantum field theory \cite{St}. There was
no indication, however, that the Feynman  $u$ parameter was taken to have any
independent physical meaning. Our framework discussed above certainly asks for the
$\sigma$ parameter to be taken as a truly important physics parameter beyond the
$\tau/m$ limit. And one should take special caution against thinking about it
too much as a quantity analogous to any `time' variable. For example, Ref.\cite{030}
illustrates it has a close connection to a scaling parameter in the full quantum
relativity. The challenge to fully appreciate the $\sigma$ variable is beyond
us here, but sure a main target of the research program.

In this letter, we introduce the Poincar\'e-Snyder relativity and Snyder
relativity as relativities in between the well known Galilean and Einstein
cases and the quantum relativity --- the relativity for `quantum space-time'.
We illustrate, using the symmetry group geometric quantization framework, how
the Poincar\'e-Snyder relativity may be providing a stronger framework for the
description of the usual relativistic quantum mechanics, from the perspective
of the which the formulation under Einstein relativity is sort of an incomplete
description. The extra `evolution' parameter $\sigma$ have been actually used
in various limiting form by earlier authors. Our Poincar\'e-Snyder relativity
provides a formulation for thinking about the $\sigma$ variable in more serious
manner, on a similar footing as the space and time variables. We will report
further on the physics of $\sigma$-evolutions in future publications.

\section{appendix : Invariant Vector Fields on SO(1,3) Group Manifold}
We give in this appendix some details of our results on
invariant vector fields on the $SO(1,3)$ group manifold.
Starting with a generic group element
$\Lambda(\omega) =e^{\frac{-i}{2}\omega^{\mu\nu} J_{\mu\nu}}$ in terms of
the generators $J_{\mu\nu}$ in the standard form, we rewrite it in terms
of two commuting sets of generators for separate $SU(2)$ groups as
$\Lambda(\omega) =   e^{-i({\omega_+}^iJ^+_i)} e^{-i({\omega_-}^iJ^-_i)}$,
where $J^\pm_i=\frac{1}{2}\left( \frac{1}{2}{\epsilon_i}^{jk} J_{jk} \pm i J_{0i} \right)$
respecting $[J^\pm_i,J^\pm_j]=i{\epsilon_{ij}}^{k}J^\pm_k$, and
${\omega_\pm}^i= \frac{1}{2}{\epsilon^i}_{jk}\omega^{jk}\mp i \omega^{0i}
\equiv \theta^i \mp i \eta^i$. The group
product may be written as
\beq
\Lambda(\omega'')=\Lambda(\omega')\Lambda(\omega)
 =e^{-i({\omega'_+}^iJ^+_i)}     e^{-i({\omega_+}^iJ^+_i)} \,
    e^{-i({\omega'_-}^iJ^-_i)}     e^{-i({\omega_-}^iJ^-_i)} \;.
\eeq
The left invariant vector fields for the $SU(2)$ group
${\bf {X}}^{\!\!\ssc L}_{\omega_\pm^{i}}$'s can be computed directly. We list the
result here dropping the + or - sign :
\beq \label{su2}
{\bf {X}}^{\!\!\ssc L}_{\omega^i}=\frac{|\omega|\cot({|\omega|}{/2})}{2}
 \frac{\partial}{\partial \omega^i}
 +\frac{2-|\omega|\cot(|\omega|/2)}{2|\omega|^2}\eta_{ij}\omega^j
  \omega^k\frac{\partial}{\partial\omega^k}-\frac{1}{2}\eta^{kh}{\epsilon_i}_{jk}
 \omega^j\frac{\partial}{\partial\omega^h} \;.
\eeq
We have then from the relations
${\bf {\widetilde{X}}}^{\!\!\ssc L}_{\theta^i}={\bf {X}}^{\!\!\ssc L}_{\omega_+i}+{\bf {X}}^{\!\!\ssc L}_{\omega_-^i}$ and
${\bf {\widetilde{X}}}^{\!\!\ssc L}_{\eta^i}=-i\left({\bf {X}}^{\!\!\ssc L}_{\omega_+i}-{\bf {X}}^{\!\!\ssc L}_{\omega_-^i}\right)$
\beqa
{\bf {\widetilde{X}}}^{\!\!\ssc L}_{\theta^{i}}  &=&
\frac{A}{2}\frac{\partial}{\partial\theta^{i}} + B\frac{\partial}{\partial\eta^{i}}
+\frac{1}{2} (C\theta_i-D\eta_i) \left( \theta^k \frac{\partial}{\partial\theta^k}
 + \eta^k  \frac{\partial}{\partial\eta^k} \right)
     +\frac{1}{2} (C\eta_i+D\theta_i) \left( \theta^k \frac{\partial}{\partial\eta^k}
   - \eta^k  \frac{\partial}{\partial\theta^k} \right)
\nonumber \\ && 
 -\frac{1}{2} {\epsilon_{ij}}^k  \left( \theta^j\frac{\partial}{\partial\theta^k}
           +\eta^j\frac{\partial}{\partial\eta^k} \right)  \;,
\nonumber \\
{\bf {\widetilde{X}}}^{\!\!\ssc L}_{\eta^{i}} &=&
\frac{A}{2}\frac{\partial}{\partial\eta^{i}} - B\frac{\partial}{\partial\theta^{i}}
-\frac{1}{2} (C\eta_i+D\theta_i) \left( \theta^k \frac{\partial}{\partial\theta^k}
 + \eta^k  \frac{\partial}{\partial\eta^k} \right)
 +\frac{1}{2} (C\theta_i-D\eta_i) \left( \theta^k \frac{\partial}{\partial\eta^k}
   - \eta^k  \frac{\partial}{\partial\theta^k} \right)
\nonumber \\ &&
  -\frac{1}{2}  {\epsilon_{ij}}^{k}  \left( \theta^j\frac{\partial}{\partial\eta^k}
     - \eta^j  \frac{\partial}{\partial\theta^k}   \right)        \;,
\eeqa
[we mark $SO(1,3)$ vector fields with 
${\bf {\widetilde{X}}}^{\!\cdot\cdot}_{\cdot\cdot}$ instead of
just ${\bf {{X}}}^{\!\cdot\cdot}_{\cdot\cdot}$, following the main text],  where
\beqa
A(\omega)&=&
\frac{\alpha\sin{\alpha} +\beta\sinh\beta}{\cosh\beta-\cos\alpha}\;,
%
\qquad\qquad\qquad
B(\omega)   \;=\;
\frac{1}{2}\frac{\beta\sin{\alpha} -\alpha\sinh\beta}{\cosh\beta-\cos\alpha}\;,
\nonumber \\
C(\omega)&=&
\frac{2(\alpha^2-\beta^2)   (      \cos\alpha-     \cosh\beta)
                +  (\alpha^2+\beta^2)   (\alpha\sin\alpha-\beta\sinh\beta)}
                 { (\alpha^2+\beta^2)^2 (\cos\alpha-\cosh\beta)} \;,
\nonumber \\
D(\omega)&=&
-\frac{4\alpha\beta(\cos\alpha-\cosh\beta)+(\alpha^2+\beta^2)(\beta\sin\alpha 
 +\alpha\sinh\beta)}
{\left(\alpha^2+\beta^2\right)^2(\cos\alpha-\cosh\beta)} \;,
\nonumber \\ &&
\alpha=\frac{1}{\sqrt{2}}\, \sqrt{  \frac{1}{2}\omega_{\mu\nu}\omega^{\mu\nu}
      +\sqrt{  \left(  \frac{1}{2} \omega_{\mu\nu} \omega^{\mu\nu} \right)^2
      +\left( \frac{1}{4} \epsilon_{\mu\nu\rho\sigma} \omega^{\mu\nu} \omega^{\rho\sigma}
   \right)^2  }}  \;,
\nonumber \\ &&
\beta=\frac{1}{\sqrt{2}}\, \sqrt{ - \frac{1}{2}\omega_{\mu\nu}\omega^{\mu\nu}
      +\sqrt{  \left(  \frac{1}{2} \omega_{\mu\nu} \omega^{\mu\nu} \right)^2
       +\left( \frac{1}{4} \epsilon_{\mu\nu\rho\sigma} \omega^{\mu\nu} \omega^{\rho\sigma}
       \right)^2  }} \;.
\eeqa
The above expressions can be written in the recombined form as
\beqa
{\bf \widetilde{X}}^{\!\!\ssc L}_{\omega^{\mu\nu}}             &=&
\frac{1}{2}A(\omega)\, \frac{\partial}{\partial\omega^{\mu\nu}}
 -\frac{1}{2}B(\omega)\,\epsilon_{\mu\nu\rho\sigma} \frac{\partial}{\partial\omega_{\rho\sigma}}
 +\frac{1}{2}C_{\mu\nu}(\omega)\omega^{\rho\sigma}\frac{\partial}{\partial\omega^{\rho\sigma}}
\nonumber \\ &&
    -\frac{1}{8}\epsilon_{\mu\nu\tau\lambda}C^{\tau\lambda}(\omega)
     \epsilon^{\alpha\beta\rho\sigma}
\omega_{\alpha\beta} \frac{\partial}{\partial\omega^{\rho\sigma}}
+\frac{1}{2}\omega^{\rho\sigma}
\left( \eta_{\mu\rho}\frac{\partial\quad}{\partial\omega^{\nu\sigma}}
+ \eta_{\nu\sigma}\frac{\partial\quad}{\partial\omega^{\mu\rho}}      \right)  \;,
\eeqa
where
\beqa
{C}_{\mu\nu}(\omega)&=&
\frac{1}{2}\left[C(\omega)\omega_{\mu\nu}
        -\frac{1}{2}D(\omega)\epsilon_{\mu\nu\rho\sigma}\omega^{\rho\sigma}\right]\;.
\eeqa
Similarly, we have
\beqa
{\bf \widetilde{X}}^{\!\!\ssc R}_{\omega^{\mu\nu}}             &=&
\frac{1}{2}A(\omega)\, \frac{\partial}{\partial\omega^{\mu\nu}}
 -\frac{1}{2}B(\omega)\,\epsilon_{\mu\nu\rho\sigma} \frac{\partial}{\partial\omega_{\rho\sigma}}
 +\frac{1}{2}C_{\mu\nu}(\omega)\omega^{\rho\sigma}\frac{\partial}{\partial\omega^{\rho\sigma}}
\nonumber \\ &&
    -\frac{1}{8}\epsilon_{\mu\nu\tau\lambda}C^{\tau\lambda}(\omega)
     \epsilon^{\alpha\beta\rho\sigma}
\omega_{\alpha\beta} \frac{\partial}{\partial\omega^{\rho\sigma}}
-\frac{1}{2}\omega^{\rho\sigma}
\left( \eta_{\mu\rho}\frac{\partial\quad}{\partial\omega^{\nu\sigma}}
+ \eta_{\nu\sigma}\frac{\partial\quad}{\partial\omega^{\mu\rho}}      \right)  \;.
\eeqa

The left- and right-invariant vector fields for the Lorentz group are available
in the literature, for example in Ref.\cite{ap}, typically not in the form 
directly with $\omega^{\mu\nu}$ as coordinate parameters. Indeed, we do not find
even expression (\ref{su2}) in the literature. For instance, vector fields 
are given in Ref.\cite{ap} with a prior splitting between the Lorentz 
boosts and rotations, each in terms of parameters that constitute a 3D vector.
\footnote{The actual vectorial parameter for the the rotation part for example
is essentially a sine function of the $|\omega/2|$. In particular, the explicit
form as given there has an extra factor of 2 coming up in the commutator, which
is a result of a $1/2$ factor when each of the parameters is matched to 
$\omega^{ij}$ in the infinitesimal limit effectively rescaling the generators. 
} 
We find the form as presented here more appealing, 
and hope that it can be generalized to $SO(p,q)$. The vector fields 
actually have little role to play in the quantization procedure as shown above. As for
the expressions of the other $G(1,3)$ invariant vector fields, we use the Lorentz
transformation matrix $\Lambda^\mu_{\!\nu}$ instead. We consider the expressions
illustrating more transparent physics. 

\bigskip
\bigskip

\noindent{\em Acknowledgements :-\ }
Otto Kong would like to thank the Korea Institute for Advanced Study for
hospitality when quite a part of an early version of the work was performed.
This work is partially supported by the research Grant No.
96-2112-M-008-007-MY3 from the NSC of Taiwan.


\newpage

\begin{table}[t]
 \caption{Physics of the Relativity Deformations Summarized: The first
column shows the familiar Galiean to Einstein case.
Having an invariant speed $c$ change the velocity from a unconstrained three-vector $v^i$
to a constrained four-vector $u^\mu$, matched to a 4D arena with an extra coordinate $ct$.
Mathematically, the zero commutator between Galilean boosts is deformed to a rotation.
The latter two columns similarly summarized the physics of the two further steps 
towards the quantum relativity of Ref.\cite{030}. For example (2nd column), having an 
invariant magnitude $\kappa c$ for $p^\mu=\frac{dx^\mu}{d\sigma}$ change the unconstrained 
four-vector to a constrained five-vector $\pi^{\ssc A}$, to be matched to a 5D arena 
with an extra coordinate $\kappa c \sigma$. The orginally zero commutator between two 
momentum boosts, to be described before imposing the constraint as
 $\Delta x^\mu = p^\mu \sigma$ on 4D space-time, is deformed to a Lorentz transformation.
The introduction of the $\sigma$ parameter/coordinate and the momentum boosts before
the deformation are the key features behind the Poincar\'e-Snyder relativity introduced
in this letter.
}
\begin{center}
\begin{tabular}{|ccc|}    \hline\hline
$\Delta x^i(t) = {v^i} \cdot t$   &   $\Delta x^\mu(\sigma) =
{p^\mu} \cdot \sigma$
 &  $\Delta x^{\ssc A}(\rho) = {z^{\ssc A}} \cdot \rho$  \\
\hline
{$|v^i|\leq c$}           &   {$\sqrt{-\eta_{\mu\!\nu} p^\mu p^\nu}\leq \kappa\,c$ }      
     &    {$| z^{\ssc A} |\leq \ell $} \\
$ \eta_{ij} v^i v^j = c^2 \left(1-\frac{1}{\gamma^2}\right) $
 &   $\eta_{\mu\!\nu} p^\mu p^\nu = -\kappa^2 c^2  \left(1-\frac{1}{\Gamma^2}\right)$
      & $\eta_{\ssc \!A\!B} z^{\!\ssc A} z^{\!\ssc B}= \ell^2 \left(1+\frac{1}{G^2}\right)$     \\
\hline
$M_{i{\ssc 0}}\equiv K_i \sim P_i$                  &   $J_{{\!\ssc 4}\mu}\equiv O_\mu \sim P_\mu $
      &   $J_{\ssc \!5\!A}\equiv O'_{\ssc \!A} \sim P_{\ssc \!A} $    \\
{$[K_i, K_j]  \longrightarrow -\, M_{ij}$}   &   {$[O_{\!\mu}, O_{\!\nu}]  \longrightarrow \, M_{\mu\nu}$}
     &  {$[O_{\!\ssc A}^{\prime} , O_{\!\ssc B}^{\prime} ]  \longrightarrow  \, J_{\ssc \!A\!B}$} \\
\hline $\vec{u}^4=\frac{\gamma}{c}(c , v^i)  $ &
$\vec{\pi}^5=\frac{\Gamma}{\kappa\,c}(p^\mu , \kappa\,c) $  &
        $\vec{X}^6=\frac{G}{\ell}(z^{\ssc A} , \ell) $\\
{$-\eta_{\mu\nu} u^\mu u^\nu = 1$}     &   {$\eta_{\ssc A\!B} \pi^{\ssc A} \pi^{\ssc B} = 1 $}
 &   {$\eta_{\ssc \mathcal M\mathcal N} X^{\!\ssc \mathcal M} X^{\!\ssc \mathcal N} = 1$}  \\
{$I\!\!R^3 \!\! \rightarrow SO(1,3)/SO(3)$}       &   {$M^4  \!\!\rightarrow SO(1,4)/SO(1,3)$} &
   {$M^5  \!\!\rightarrow SO(2,4)/SO(2,3)$}
\\
\hline\hline
\end{tabular}
\end{center}
\end{table}


\begin{table}[t]
 \caption{The various relativities -- matching the generators :
The table matches out the genarators for the various relativity symmetries 
from a pure mathematical point of view. Note as algebras, the mathematical
structures of translations (denoted by $P_{.}$) or the boosts (denoted by 
$K_{.}$ and   $K_{.}'$ -- the so-called Lorentz boosts not includedas they are
really space-time rotations) in relation to rotations $J_{..}$ are the same. 
Algebraically, translation and boost generators are distinguished only by
the commutation with the Hamiltonian ($H$ and   $H'$). Successive contractions
retrieve $G(1,3)$ and $ISO(1,4)$ from $SO(2,4)$, similar to the more familiar
$G(3)$ and $ISO(1,3)$	
from $SO(1,4)$. In the physics picture under discussion, however, $SO(1,4)$
part of our so-called Snyder relativity $ISO(1,4)$ is {\it different} from
the usual de-Sitter $SO(1,4)$ contracting to $ISO(1,3)$. We consider simply 
keeping only the  $P_{\mu}$ and  $J_{\mu\nu}$ generators to reduce from
our Poincar'e-Snyder $G(1,3)$ to the Einstein $ISO(1,3)$.
}
\begin{center}
\begin{tabular}{|c|ccccc|}    \hline\hline
Relativity	& Quantum	& Snyder	& Poincar\'e-Snyder	& Einstein		& Galilean \\
Symmetry	& $SO(2,4)$ & $ISO(1,4)$	& $G(1,3)$		& $ISO(1,3)$	&  $G(3)$\\
\hline
Arena		&`AdS$_5$'	& $M^5$	& $M^4$ (with $\sigma$)	& $M^4$	& $I\!\!R^3$ (with $t$)\\
\hline
\hline
		& $J_{ij}$  	& $J_{ij}$	& $J_{ij}$  	& $J_{ij}$	& $J_{ij}$\\
SO(1,4)	& $J_{i\ssc 0}$		& $J_{i\ssc 0}$	& $J_{i\ssc 0}$        & $J_{i\ssc 0}$	& $K_{i}$ \\
part		& $J_{\ssc 40}$	& $J_{\ssc 40}$	& $K_{\ssc 0}^{\prime}$	& $P_{\ssc 0}$	&  $H$ \\
		& $J_{{\ssc 4}i}$	& $J_{{\ssc 4}i}$	& $K_i^{\prime}$		& $P_i$ 	&  $P_i$ \\
\hline
		& $J_{\ssc 50}$	&  $P_{\ssc 0}$	&  $P_{\ssc 0}$	& 	&\\	
		& $J_{{\ssc 5}i}$  &  $P_i$		&  $P_i$		&   	&\\
		& $J_{\ssc 54}$	&  $P_{\ssc 4}$ 	&  $H^{\prime}$	&	&\\
\hline\hline
\end{tabular}
\end{center}
\end{table}

\end{document}